\documentclass[graybox]{svmult}

\usepackage{geometry}
\usepackage{makeidx}         % allows index generation
\usepackage{url}
\usepackage{graphicx}        % standard LaTeX graphics tool
\usepackage{multicol}        % used for the two-column index
\usepackage[bottom]{footmisc}% places footnotes at page bottom

\usepackage{newtxtext}       % 
\usepackage[varvw]{newtxmath}       % selects Times Roman as basic font

\usepackage[numbers]{natbib}
\usepackage[version=4]{mhchem}
\usepackage{csquotes}

\usepackage{xcolor, soul}
\sethlcolor{green}

\makeindex             % used for the subject index
\begin{document}

\title*{Using mathematical models of heart cells to assess the safety of new pharmaceutical drugs}
\titlerunning{Drug safety simulation} % for an abbreviated version of your contribution title if the original one is too long
\author{Gary R. Mirams}
\institute{Gary R. Mirams \at Centre for Mathematical Medicine \& Biology, School of Mathematical Sciences, University Park, Nottingham, NG7 2RD, United Kingdom, \email{gary.mirams@nottingham.ac.uk}}

\maketitle

This is a preprint of the following chapter: Gary R.\ Mirams, Using mathematical models of heart cells to assess the safety of new pharmaceutical drugs, published in More UK Success Stories in Industrial Mathematics, edited by Philip J.\ Aston, 2025, volume 42 in the series \emph{Mathematics in Industry}, Springer reproduced with permission of Springer Nature Switzerland AG. The final authenticated version is available online at: \url{https://doi.org/10.1007/978-3-031-48683-8_22}
\newline
\newline
% For online index, no more than 250 words.
\abstract{
Many drugs have been withdrawn from the market worldwide, at a cost of billions of dollars, because of patient fatalities due to them unexpectedly disturbing heart rhythm. 
Even drugs for ailments as mild as hay fever have been withdrawn due to an unacceptable increase in risk of these heart rhythm disturbances. 
Consequently, the whole pharmaceutical industry expends a huge effort in checking all new drugs for any unwanted side effects on the heart.
The predominant root cause has been identified as drug molecules blocking ionic current flows in the heart.
Block of individual types of ionic currents can now be measured experimentally at an early stage of drug development, and this is the standard screening approach for a number of ion currents in many large pharmaceutical companies. 
However, clinical risk is a complex function of the degree of block of many different types of cardiac ion currents, and this is difficult to understand by looking at results of these screens independently.
By using ordinary differential equation models for the electrical activity of heart cells (electrophysiology models) we can integrate information from different types of currents, to predict the effect on whole heart cells and subsequent risk of side effects. 
The resulting simulations can provide a more accurate summary of the risk of a drug earlier in development and hence more cheaply than the pre-existing approaches. 
}

%\keywords{Mathematical Modelling, Ion Channels, Electrophysiology, Safety Pharmacology, Action Potential, hERG, Drug Binding, Pro-arrhythmic Risk}

% To comment out later
%\tableofcontents
\section{Introduction}
The contraction of your heart is both triggered and co-ordinated by a wave of electrical activity passing through each of its muscle cells. 
This wave propagates because ions of various types --- particularly sodium, potassium and calcium --- move into and out of each cell on every beat.
The ions move through thousands of protein complexes that are in each cell's membrane. 
Some of these protein complexes passively allow specific types of ions to flow along their electrochemical gradient (this would be simply from high to low concentration but there is also a contribution from the electric field), these are called \emph{ion channels}.
Other complexes are \emph{ion pumps}, which use energy to actively move an ion against its gradient; and another type are \emph{ion exchangers}, which move one type of ion against its gradient using the energy from the flow of another type down its gradient.
These channels, pumps and exchangers are encoded by particular genes, and typically we consider 10--20 genes and their associated types of current carriers to be the most important in explaining the heart's electrical activity.
Many of the currents are selective for just particular ions, and they respond to the trans-membrane voltage of the cell in different ways (opening or closing, rapidly or slowly, at different voltages).

Some of the heart's ion channels are particularly prone to being blocked by many pharmaceutical drug compounds. 
Sometimes this is the desired action of a drug, to treat disturbances to healthy heart rhythm which are known as \emph{arrhythmias}.
But unfortunately many drugs (even for diseases unrelated to the heart) also block cardiac ion channels as a side effect; this can have serious consequences in terms of increasing the risk of patients suffering fatal arrhythmias.
As a result large amounts of time and money are now spent during the drug development process to test whether potential drug compounds are likely to have these pro-arrhythmic side effects on the heart. 

In the first instance, pharmaceutical companies can check whether new compounds interact with cardiac ion channels in automated experiments for one type of ion channel at a time.
This `screening' effort has been relatively successful; very few compounds reach the market and then have unexpected effects on the heart.
Regulators of the pharmaceutical industry, such as the USA's Food \& Drug Administration, have become more concerned that we may be discarding potentially good drug compounds that have small effects on the heart which would not actually be pro-arrhythmic.
So in recent years we have been working with pharmaceutical companies and international regulators to run electrophysiology simulations using mathematical models to improve and refine the assessment of pro-arrhythmic risk of novel drug compounds.

\newpage
\section{Modelling the cardiac action potential}

\begin{figure}[hbt]
    \centering
    \includegraphics[width=0.9\textwidth]{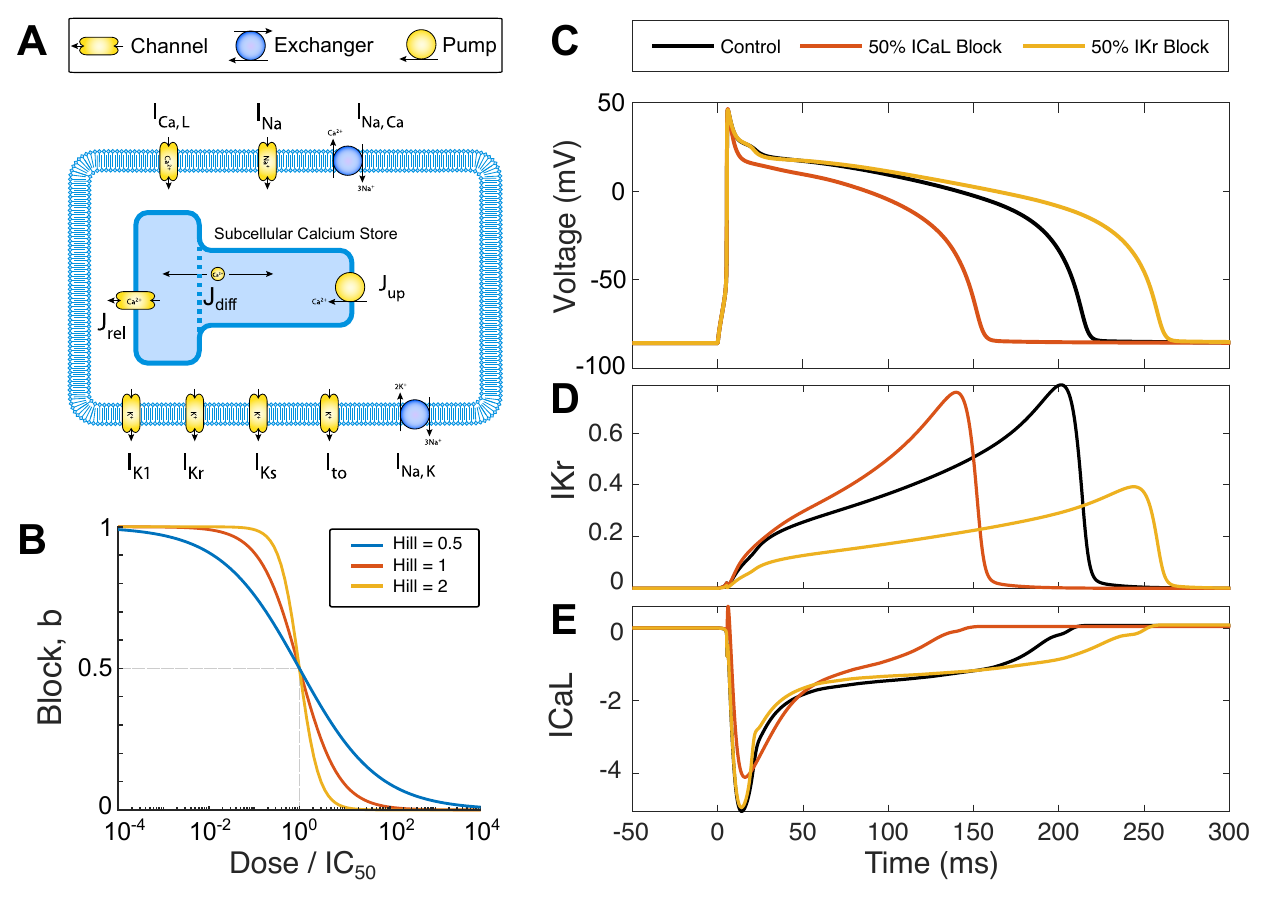}
    % Generated using the scripts in the figures folder to make ingredients for figshare fig.
    \caption{Models for the electrical activity of cardiac cells and drug block of ion currents.
    \textbf{(A)} A schematic of a cardiac electrophysiology model showing the predominant ionic currents and the sub-cellular calcium store (sarcoplasmic reticulum), adapted from \protect\url{www.cellml.org}. 
    The voltage across the cell membrane will change in response to the sum of all the currents crossing the outer membrane, following Eq.~(\ref{eqn:dVdt}).
    \textbf{(B)} The Hill equation shown with the ratio of dose and IC\textsubscript{50} on the $x$-axis, varying Hill coefficients ($n$ in Eq.~(\ref{eq:Hill})) shown. A simulation using the Shannon 2004 rabbit ventricle model (\protect\url{https://bit.ly/shannonrabbit}) showing the effects of drug block of I\textsubscript{Kr} or I\textsubscript{CaL} by 50\% on:
    \textbf{(C)} the membrane voltage / action potential waveform;
    \textbf{(D)} IKr, an outward repolarising current; and
    \textbf{(E)} ICaL, an inward current (currents shown in $\mu$A/cm$^2$).}
    \label{fig:models}
\end{figure}

The fundamental model for electrical activity of cells is relatively simple: the voltage across the cell membrane ($V$) changes as a result of ionic currents flowing across the membrane.
In ordinary differential equation (ODE) form, this reads
\begin{equation}
\frac{\mathrm{d}V}{\mathrm{d}t} = - \frac{1}{C_m} \sum_i^N I_i,
\label{eqn:dVdt}
\end{equation}
where $C_m$ is the cell capacitance, and there are $N$ different types of ionic membrane currents, each denoted by $I_i$.
We show an example model in Fig.~\ref{fig:models}A.
This becomes an interesting and complex feedback model because the currents $I_i$ themselves evolve according to more ODEs that are nonlinear functions of $V$.

\newpage
The simplest ion channel current models typically look like
\begin{equation}
I_i = g_i \cdot \mathcal{P}_i(V) \cdot (V-E_i),
\label{eq:ohmic}
\end{equation}
where $\mathcal{P}_i$ is the open probability of the channels of a given type $i$, $g_i$ is a parameter representing the maximal current that can flow when these channels are open (considered proportional to the density of channels in the membrane), and $E_i$ is the Nernst/reversal potential for this type of ion.
The open probability $\mathcal{P}_i$ is a function of $V$, and this is usually represented with further sub-systems of ODEs, describing how rapidly this particular current's channels open and close in response to the applied voltage ($V$).

\section{Modelling drug action on ion channels}
We can modify Eq.~(\ref{eq:ohmic}) to include a scaling factor, $b$, that represents the proportion of current remaining after block with a drug at concentration $[\text{D}]$
\begin{equation}
I = g \cdot b([\text{D}]) \cdot \mathcal{P}(V) \cdot (V-E). \label{eq:bgatecurrent}
\end{equation}
To derive an expression for $b$ as a function of drug concentration we consider the following reversible chemical reaction: $n$ drug compounds (D) bind to one ion channel (C) to produce a complex (CD$_n$) where the ion channel is blocked, at forward rate $k$\textsubscript{on} and backward rate $k$\textsubscript{off}:
\begin{equation}
\ce{C + nD  <=>[$k_{\text{on}}$][$k_{\text{off}}$] CD_n}
\end{equation}
Assuming mass-action kinetics (reaction rates are directly proportional to the product of their reactant concentrations) allows us to formulate an ODE for C:
\begin{equation}
\frac{\mathrm{d}[\text{C}]}{\mathrm{d}t} = -k_{\text{on}}[\text{C}][\text{D}]^n  + k_{\text{off}} [\text{CD}_n],
\label{eq:mass_act}
\end{equation}
where square brackets denote concentrations.
We can divide through by the total number of channels (a constant, $[\text{C}]+[\text{CD}_n]$) to work with $b$, defined as the proportion of un-blocked channels as above.
At steady state we get
\begin{equation}
b_\infty([\text{D}]) = \frac{\frac{k_{\text{off}}}{k_{\text{on}}}}{[\text{D}]^n + \frac{k_{\text{off}}}{k_{\text{on}}}}.
\end{equation}
At 50\% block we will have $[\text{D}]^n = \frac{k_{\text{off}}}{k_{\text{on}}}$; the concentration at which this happens is called the `50\% Inhibitory Concentration' or $[\text{IC}_{50}]$, so that $[\text{IC}_{50}]^n = \frac{k_{\text{off}}}{k_{\text{on}}}$. 
We can substitute this in to get
\begin{equation}
b_\infty([\text{D}]) = \frac{[\text{IC}_{50}]^n}{[\text{D}]^n + [\text{IC}_{50}]^n} = \frac{1}{1 + \left( \frac{[\text{D}]}{[\text{IC}\textsubscript{50}]}\right)^n}.
\label{eq:Hill}
\end{equation}
This equation is known as the Hill equation. We can substitute it in place of $b$ within Eq.~(\ref{eq:bgatecurrent}) to simulate the steady-state block of a current as a function of drug concentration, which is then parameterised by the IC\textsubscript{50} and $n$ which is known as the `Hill coefficient'.
In Fig.~\ref{fig:models}B we show plots of Eq.~(\ref{eq:Hill}) for three values of the Hill coefficient $n$.

Fig.~\ref{fig:models}C shows how the cell's membrane voltage changes in response to $b_\infty = 0.5$, i.e.\ 50\% block, for either I\textsubscript{Kr} or I\textsubscript{CaL}, and the coupled changes to currents in Fig.~\ref{fig:models}D\&E.
In reality, different compounds will have different affinities for many different channels, and the models as described above allow us to predict the whole-cell effect on electrophysiology.

\section{Evaluating model predictions}

In \citet{mirams2011simulation} we used the approach outlined above for multiple ion channels (I\textsubscript{Kr}, I\textsubscript{CaL}, I\textsubscript{Na}), and predicted the resulting action potential duration (length of the wave in Fig.~\ref{fig:models}C). 
We showed how classifying drug risk based on this simulation output was more predictive of clinical risk of arrhythmia than simply considering the I\textsubscript{Kr} IC\textsubscript{50} or safety margin (concentration of drug reached in clinical dosing / IC\textsubscript{50}), which had been the predominant strategy up to that point.

Another study tested model predictions for the effects of hundreds of compounds on rabbit heart tissue that had been measured over many years as part of drug development at GlaxoSmithKline Plc.\ \citep{Beattie2013}. 
This study showed that simulations could reach useful levels of predictive power across a whole portfolio of compounds under development within the pharmaceutical industry, and highlighted the accuracy of different pharmaceutical screening technologies.

\begin{figure}[hbt]
    \centering
    \includegraphics[width=\textwidth]{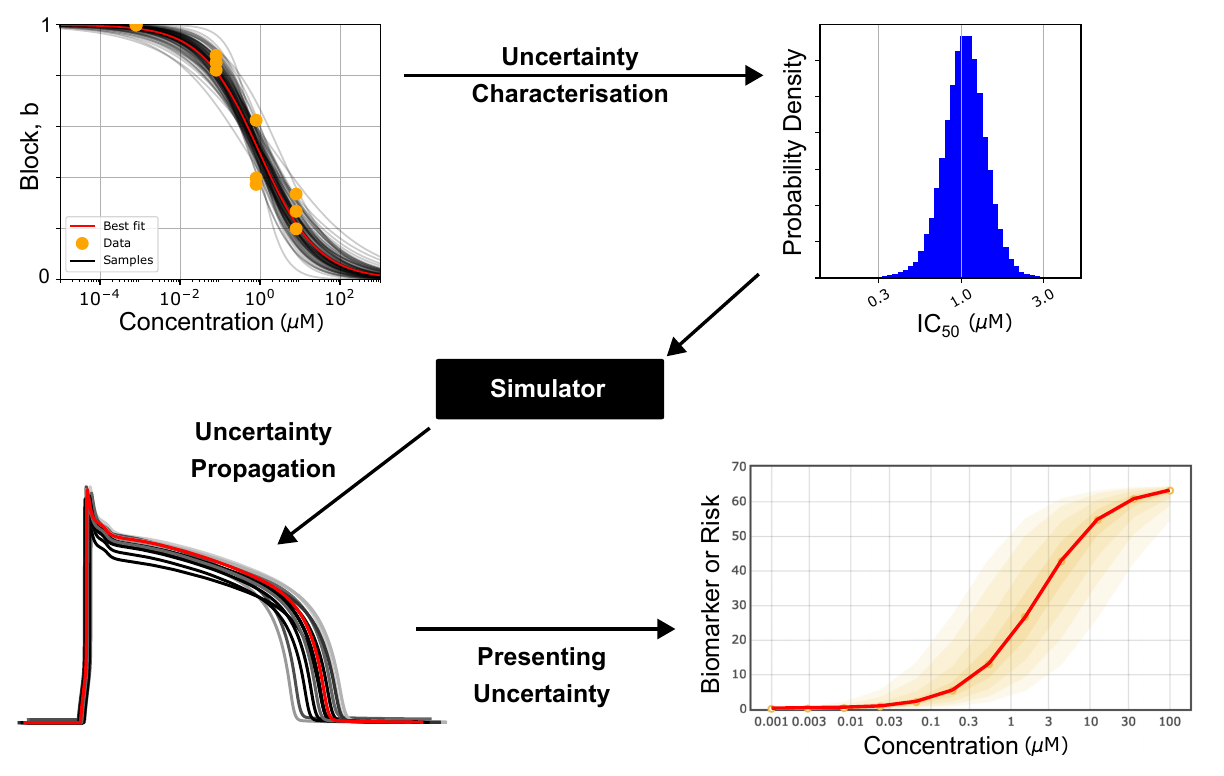}
    % Generated using the scripts in the figures folder to make ingredients for figshare fig.
    \caption{Uncertainty Quantification for simulations of drug action on electrophysiology. 
    For each ion current that is screened some concentration/block data points are measured. 
    There is resulting uncertainty in the underlying compound properties such as IC\textsubscript{50}, that we can estimate with an `uncertainty characterisation' procedure such as Bayesian inference for the parameters of interest (IC\textsubscript{50} shown here in top right, with corresponding sampled Hill curves in the top left).
    The simulator then takes as an input a probability distribution, rather than a single number, and produces a corresponding distribution of outputs.
    The end-user is then presented with a visualisation of the probability that simulation outputs (such as clinical risk) will be in certain ranges, shown here as a fan chart.}
    \label{fig:uq}
\end{figure}

Because there is uncertainty in the experimental data that the simulations are based upon it is important to consider the consequences of this on model outputs.
As shown in Fig.~\ref{fig:uq} and discussed in its caption, the first part of this task is `uncertainty characterisation' --- establishing the uncertainty or variability in the model inputs \citep{johnstone2016hierarchical}.
The second part of the task is `uncertainty propagation', mapping the input uncertainty through a simulator to see how this uncertainty affects simulation outputs. This is especially important when predicting clinical risk classes \citep{li_assessment_2019} that may decide whether a drug progresses further in development or is approved without restrictions.

These validation studies and a consideration of uncertainty in the simulation results provided enough help with decision making for the pharmaceutical industry and pharmaceutical regulators to pursue the impact discussed below.

\section{Impact}

A web portal is publicly available (\url{https://cardiac.nottingham.ac.uk}) where anyone can register for an account and run cardiac safety simulations by entering data on the degree to which compounds block specific ion channels \citep{williams2015web}.
As of 2021, this tool has performed over 7000 simulations for over 350 users from 147 companies including Contract Research Organisations, small pharmaceutical companies and 14 of the top 20 largest global pharmaceutical companies; and over 70 educational or research establishments.

The simulator has also been deployed inside company firewalls at GlaxoSmithKline and Roche and is used by Safety Pharmacology teams at both companies for routine safety testing and compound profiling. 
The simulation results are used to propose concentrations for further tests, check for alignment of later test results with expectations from ion channel screening, and even to replace some animal tissue experiments completely.

The Comprehensive in-vitro Pro-arrhythmia Assay (CiPA) is a worldwide effort of the pharmaceutical industry, regulators and academia to improve pro-arrhythmic risk assessment for new drugs (\url{https://www.cipaproject.org}), and it has included the simulation approaches discussed above.
Our work on uncertainty quantification with the US Food \& Drug Administration (FDA) led to changes in the proposed clinical risk output from CiPA simulations to make it more robust. 
The new risk marker was tested in a blinded validation study \citep{li_assessment_2019}, found to provide excellent predictions, and is now the FDA’s primary pre-clinical risk marker to assess the pro-arrhythmic risk of all new pharmaceutical drugs. 
Pharmaceutical safety testing guidelines are aligned worldwide by the International Council for Harmonisation of Technical Requirements for Pharmaceuticals for Human Use (ICH), and they recently completed a new Question \& Answer document to provide guidelines on how to use simulated pro-arrhythmic risk predictions in line with the CiPA approach.
Pharmaceutical companies and regulators worldwide can now use the simulation approaches discussed above to assess the safety of new compounds.

\begin{acknowledgement}
The author gratefully acknowledges support from the Wellcome Trust (grant number 212203/Z/18/Z). 
He would also like to acknowledge helpful discussions with his research team based at the University of Nottingham, collaborators and members of the Comprehensive in-vitro Proarrhythmia Assay (CiPA) Steering Committee and working groups.
\end{acknowledgement}

\bibliographystyle{spbasic}
\bibliography{refs.bib}

\end{document}